# Development of submillimeter wave source Gyrotron FU Series for plasma diagnostics

**T. Idehara**, I. Ogawa, S. Mitsudo, M. Kamada, O. Watanabe and La Agusu

*Research Center for Development of Far Infrared Region, University of Fukui*

*Bunkyo 3-9-1, Fukui –shi 910-8507, Japan*

**Abstract**: Our gyrotrons developed in Fukui University, Research Center for Development of Far Infrared Region (FIR FU) are high frequency, medium power gyrotrons as millimeter to submillimeter wave radiation sources for application to new far-infrared technologies including plasma diagnostics. We have already developed Gyrotron FU Series which consists of 8 gyrotrons. The gyrotron series has achieved frequency tuneability in wide range (from 38 GHz to 889 GHz). The highest frequency is corresponding to a wavelength of 337 $\mu$ m. This is a current record for high frequency operation of gyrotron. Recently, we have developed a high harmonic gyrotron with an axis-encircling electron beam and a THz gyrotron with a pulse magnet. In this presentation, the present status of Gyrotron FU Series is described.

## 1. Introduction

The development of gyrotrons are proceeding in two directions. One is the development of high power, millimeter wave gyrotrons as the power sources for electron cyclotron heating of plasmas, electron cyclotron current drive of tokamaks and for ceramic sintering. The second direction is the development of high frequency, medium power gyrotrons as millimeter to submillimeter wave sources for plasma scattering measurements, ESR experiments and so on. Gyrotrons developed in FIR FU 'Gyrotron FU series' belong to the second group.

In order to develop high frequency gyrotrons, we should use high magnetic field which is generated by superconducting magnet[1] and/or operation at high harmonic of electron cyclotron frequency[2]. In our recent gyrotron, Gyrotron FU IVA, a 17T superconducting magnet is used. In the case, the treatment of the device is complicated, because we need to use cryogenic facilities, for example, transferring liquid helium from a container to cryostat, recovering evaporated helium to liquifying system and so on.

Recently, we have developed a high harmonic gyrotron with an axis-encircling electron beam and a permanent magnet instead of a superconducting magnet. The treatment is much simplified[3], but the field intensity is quite low comparing with a superconducting magnet. It is only around 1T. We have to use higher harmonic operations in order to increase a frequency of gyrotron. A gyrotron with an axis-encircling electron beam (so-called Large Orbit Gyrotron, LOG) is suitable for nth harmonic operation, when $TE_{n11}$ cavity mode is employed. In the case of a conventional gyrotron, higher harmonic operation is difficult to be realized. Gyrotron FU Series achieved mainly second harmonic operation. The third harmonic operation is very rare[4]. The efficiency is decreased with the harmonic number increased. On the other hand, in the case of LOG, higher harmonic operation (n=3, 4, 5) is easily excited. This is the most important advantage of LOG.

We are designing the next gyrotron with a pulse magnet. The coil of the magnet is cooled down by liquid nitrogen and the intensity of magnetic field is increased up to 31 T. The design of a gyrotron tube is also completed. We can expect to increase the frequency up to around 1.44 THz by using second harmonic operations. Construction of the whole device will be completed soon..

Development of these gyrotrons is our effort for increasing the operation frequency. In this presentation, the present status of Gyrotron FU Series is described.

## 2. The main results of Gyrotron FU series

Gyrotron FU series includes 8 gyrotrons. Each gyrotron consists of a sealed-off gyrotron tube and a superconducting magnet, except Gyrotrons FU III and FU V, which have demountable tubes. Table 1 summarizes the main results of the gyrotrons included in Gyrotron FU series.

The design of each gyrotron was carried out by computer simulations. We are using narrow cavities to get a good mode separation and then to operate the gyrotrons in many single modes on fundamentals, second and even third harmonics. Such a situation is important for development of our high frequency, harmonic gyrotrons.

Table 1 The results of Gyrotron FU series

| Name of gyrotron | Frequency range | Items which each gyrotron has achieved |
|---|---|---|
| Gyrotron FU I | 38-220 GHz | High frequency operation at 100GHz and 9 kW output power |
| Gyrotron FU E | 90-300 GHz | Radiation source for the first experiment on ESR |
| Gyrotron FU IA | 38-215 GHz | Radiation source for plasma scattering measurement of WT-3 |
| Gyrotron FU II | 70-402 GHz | Studies on mode competition and mode cooperation, Radiation source for plasma scattering measurement of CHS |
| Gyrotron FU III | 100-636 GHz | 3rd harmonic operation in single modes, Amplitude modulation, Frequency step switching |
| Gyrotron FU IV | 160-847 GHz | Frequency modulation, cw operation for high stability of amplitude and frequency |
| Gyrotron FU IVA | 160-889 GHz | Higher frequency operations by 3rd harmonics, High purity mode operation, Radiation source for ESR experiment |
| Gyrotron FU V | 186-222 GHz | cw operation for long time using a He free magnet, High stabilizations of frequency and amplitude, High purity mode operation |

## 3. Development of large orbit gyrotron (LOG)

Fig. 1 shows a schematic drawing of our recent gyrotron, LOG.

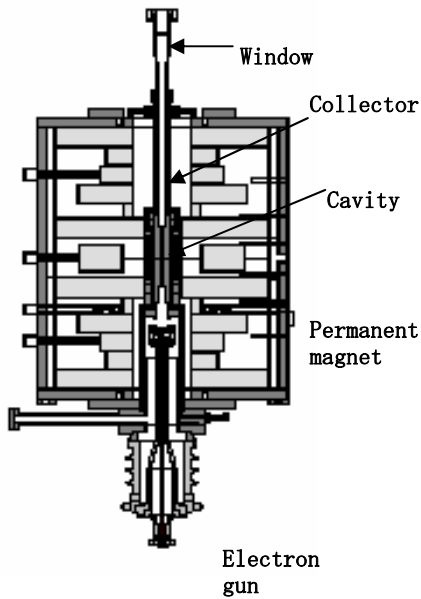

Fig.1 Schematic drawing of LOG device

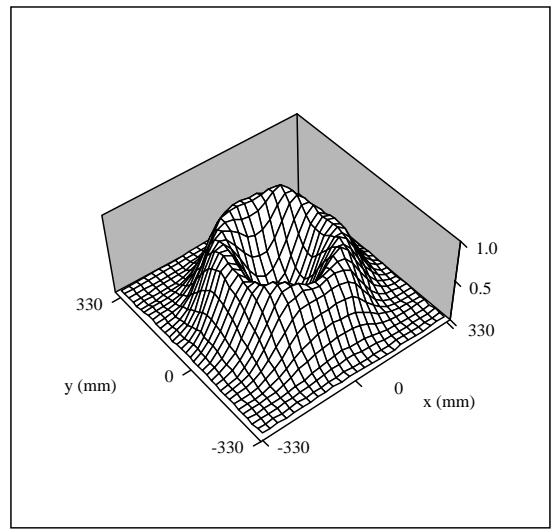

Fig. 2 Emission pattern of $TE_{311}$ mode at the third harmonic operation

It consists of a permanent magnet system and a demountable gyrotron tube. The field intensity at the center of magnet system is around 1 T. We have already tested two cavities. In the case of the first cavity (Cavity 1), which is optimized for the fourth harmonic operation using $TE_{411}$ cavity mode, we have succeeded in third, fourth and fifth harmonic operations. Output power for fourth harmonic was 0.47 kW and corresponding efficiency 0.98 percent.

For the second cavity (Cavity 2), which is optimized for the third harmonic operation using $TE_{311}$ mode. In Fig. 2, is demonstrated the emission pattern observed at 1,010 mm above the output window. It looks pure $TE_{311}$ mode. This means the mode selection in LOG is excellent. The observed output power was 2.5 kW and the efficiency 6.25 percent.

## 4. High frequency gyrotron with pulse magnet

We have just constructed a next gyrotron with a pulse magnet whose field intensity can be increased up to 23.5 T. The cavity is designed for high frequency operations at the fundamental and the second harmonic. The maximum frequency will be around 1.22 THz.

Fig. 3 shows a schematic drawing for a design of the ultra high frequency gyrotron using a pulse magnet system. It consists of a capacitor bank systems, a pulse magnet and a gyrotron tube. There are a capacitor bank downstair. A pulse magnet system can be operated by the control system which is installed upstairs. A high field pulse magnet system and a gyrotron tube are also installed upstairs.

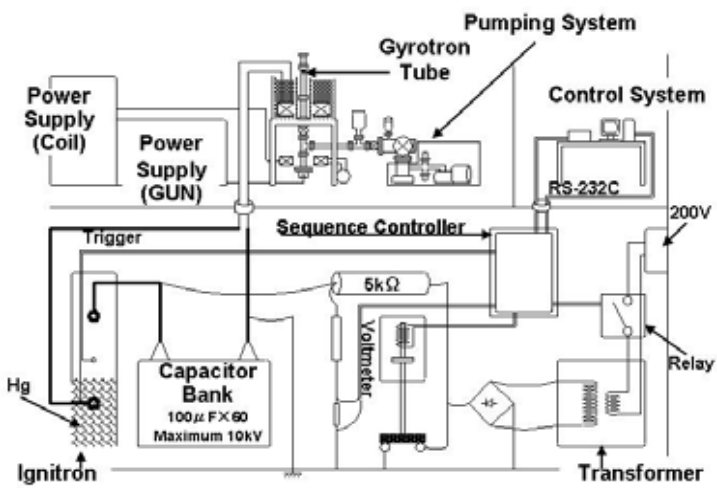
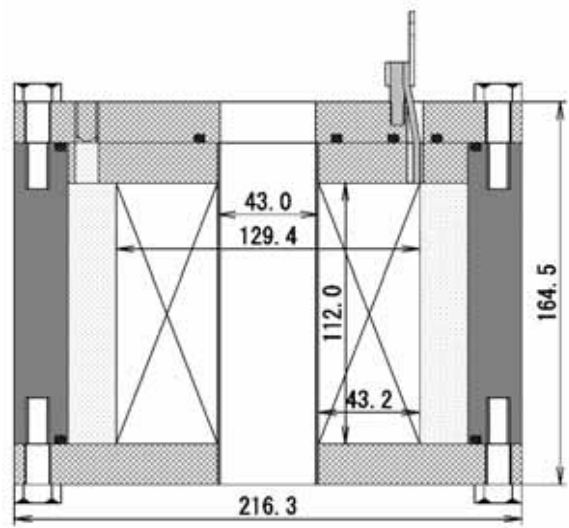

Fig. 3 A concept chart of aTHz gyrotron system         Fig. 4 A cross section of the magnet

The capacity of the bank is 6 mF and the maximum stored energy is 300 kJ at the maximum bias of 10 kV. It is switched by an ignitron triggered by a thyristor and a pulse current goes through the coil. Fig. 4 demonstrates the cross section of the magnet system. The coil is constructed by copper wire whose cross section is squire and has 320 turns. The total inductance is about 3.39 mH. A simple estimation result indicates that the pulse width of the operation of magnet system is around 14 ms and the maximum field intensity is 23.5 T. The preliminary test result of the system supports the estimation results. The uniformity in the center region of the magnet is also confirmed by the measurement. The deviation on the center axis is suppressed within 1 percent in the range of 40 mm.

The coils is installed inside the cryostat and cooled down by liquid nitrogen. Water and alumina powder surround the coil. After cooling down, water is frozen and the ice with alumina powder fixes the coil tightly.

The gyrotron tube is installed in the center air-bore of the cryostat. The tube consists of a triode magnetron injection gun (MIG), a simple cylindrical cavity and a cylindrical collector, which acts as a transmission waveguide. The radius $R_0$ and the length $L$ of the cavity are 1.5 mm and 6 mm.

*Computer simulation result of the gyrotron operation*

We have carried out the computer simulation for operation of the MIG and interaction between high frequency electromagnetic wave and beam electrons. Fig. 5 shows the result of the latter simulation, in which the parameter of beam electrons decided by the former simulation is used. In the figure in the left hand side, the starting current for each cavity mode as a function of the magnetic field intensity *B* is demonstrated for both the fundamental and the second harmonic operations. For almost all cavity modes, starting currents are less than 0.2 A. However, mode competition between neighbor modes looks severe. In particular, the second harmonic operations could not realized by the suppression effect due to the stronger excitation of the fundamental modes. In the figure in the right hand side, a computer simulation result for mode competition is demonstrated. Time evolution of several modes is shown here. It is clear that $TE_{10,7}$ mode operation at the second harmonic becomes dominant in the final stage. Therefore, this mode is one of candidates for single mode operation at the second harmonic.

In such a way, we can develop a higher frequency gyrotron with a pulse magnet. The highest frequencies expected at the fundamental and the second harmonic operations are 0.78 THz and 1.22 THz, respectively. This is the first trial for development of THz gyrotron.

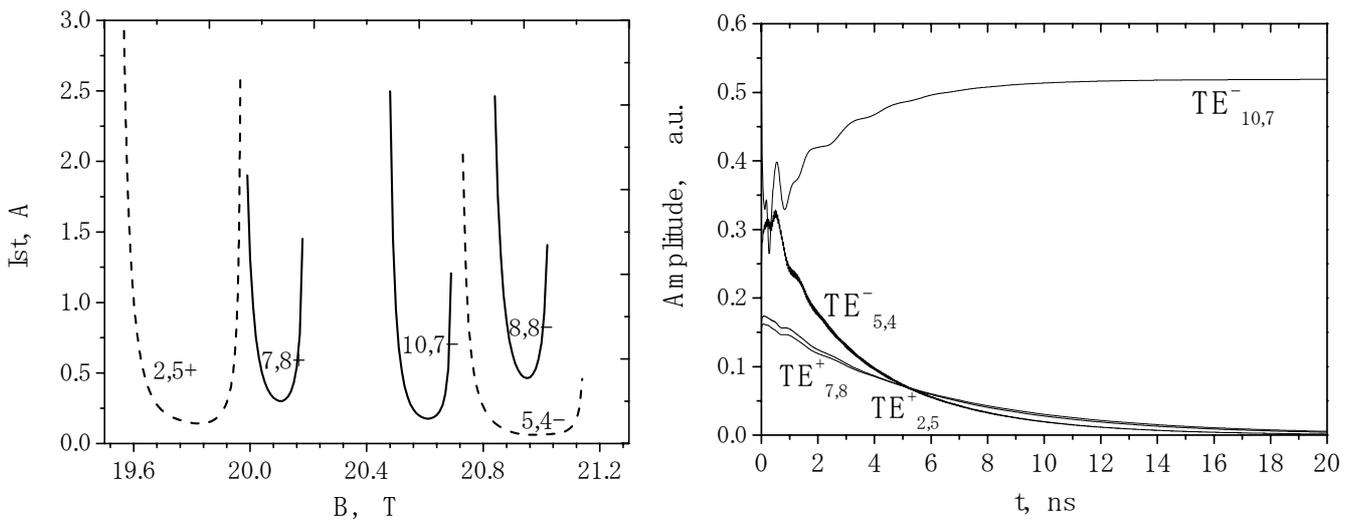

Fig. 5 (Left) Computer simulation results for starting currents *I* as functions of field intensity *B* for both fundamental (broken line) and second harmonic (solid lines) operation modes. (Right) Computer simulation results for mode competition. B=20.52 T. In final stage, $TE_{10,7}$ mode operation at the second harmonic becomes dominant.

## 4. Conclusion

For achievement of the breakthrough of 1 THz, we have already designed an ultra high frequency gyrotron with a pulse magnet system. The highest field intensity is 23.5 T and the expected maximum frequencies at the fundamental and the second harmonic operations are 0.78 THz and 1.22 THz. The construction of the whole gyrotron system has been almost completed and the operation test will begin soon.

## References

1) T.Idehara, T. Tatsukawa, S. Matsumoto, K. Kunieda, K. Hemmi and T. Kanemaki, Phys. Lett. A **132**, 43 (1988).
2) T. Idehara, T. Tatsukawa, I. Ogawa, H. Tanabe, T. Mori, S. Wada, G.F. Brand and M.H. Brennan, Phys. Fluids **B4**, 267 (1992).
3) T. Kikunaga, H. Asano, Y. Yasojima, F. Sato, T. Tsukamoto, Int. J. Electronics **79**, 655 (1995).
4) T. Idehara, T. Tatsukawa, I. Ogawa, H. Tanabe, T. Mori, S. Wada and T. Kanemaki, Appl. Phys. Lett. **56**, 1743 (1990).